\documentclass[twocolumn,aps,url]{revtex4-1}
\usepackage{graphicx}

\begin{document}
\title{Inference of Extreme Synchrony with an Entropy Measure on
a Bipartite Network} 
\author{Aki-Hiro SATO}
\affiliation{Department of Applied Mathematics and Physics, Graduate School
of Informatics, Kyoto University, Yoshida Honmachi, Sakyo-ku, Kyoto
606-8501 JAPAN}
\date{\today}
\begin{abstract}
This article proposes a method to quantify the structure of a bipartite graph
using a network entropy per link. The network entropy of a
bipartite graph with random links is calculated both numerically and
theoretically. As an application of the proposed method to analyze
collective behavior, the affairs in which participants quote and trade
in the foreign exchange market are quantified. The network entropy per
link is found to correspond to the macroeconomic situation. A finite
mixture of Gumbel distributions is used to fit the empirical
distribution for the minimum values of network entropy per link in
each week. The mixture of Gumbel distributions with parameter
estimates by segmentation procedure is verified by the Kolmogorov--Smirnov
test. The finite mixture of Gumbel distributions that extrapolate the
empirical probability of extreme events has explanatory power at
a statistically significant level.
\end{abstract}
\pacs{89.75.Hc, 89.65.Gh, 65.40.gd}
\maketitle
\section{Introduction}
The network structure of various kinds of physical and 
social systems has attracted considerable research attention. A
many-body system can be described as a network, and the 
nature of growing networks has been 
examined well~\cite{Albert:02,Miura}. 
Power-law properties can be found in the growing networks, which are
called complex networks. These properties are related to the growth of
elements and preferential attachment~\cite{Albert:02}.

A network consists of several nodes and links that connect nodes. In
the literature on the physics of socio-economic 
systems~\cite{Anna}, nodes are assumed to represent agents, goods,
and computers, while links express the relationships between
nodes~\cite{Milakovic,Lammer}. Network structure is 
perceived in many cases through the conveyance of information, knowledge,
and energy, among others.

In statistical physics, the number of combinations of possible
configurations under given energy constraints is related to
``entropy.'' Entropy is a measure that quantifies the states of
thermodynamic systems. In physical systems, entropy naturally increases
because of the thermal fluctuations on elements. Boltzmann proposed 
that entropy $S$ is computed from the possible number of ensembles
$g$ by $S = \log g$. For a system that consists of two sub-systems 
whose respective entropies are $S_1$ and $S_2$, the total
entropy $S$ is calculated as the sum of one of two sub-systems
$S_1+S_2$. This case is attributed to the possible number of ensembles
$g_1g_2$. Entropy in statistical 
physics is also related to the degree of 
complexity of a physical system. If the entropy is low (high), 
then the physical configuration is rarely (often)
realized. Energy injection or work in an observed system may be
assumed to represent rare situations. Shannon entropy is also used to
measure the uncertainty of time series~\cite{Anna:07}.

The concept of statistical--physical entropy was applied
by Bianconi~\cite{Bianconi} to measure network structure. She considered
that the complexity of a network 
is related to the number of possible configurations of nodes and links
under some constraints determined by observations. She calculated the
network entropy of an arbitrary network in several cases of
constraints. 

Researchers have used a methodology to characterize network
structure with information-theoretic
entropy~\cite{Dehmer:11,Wilhelm, Rashevsky:55, Trucco:56,
  Mowshowitz:53, Aki}. Several graph invariants such as the
number of vertices, vertex degree sequence, and extended degree
sequences have been used in the construction of entropy-based
measures~\cite{Wilhelm,Aki}.

\section{An entropy measure on a bipartite network}
\label{sec:entropy}
The number of elements in socio-economic systems is usually very large,
and several restrictions or finiteness of observations can be found.
Therefore, we need to develop a method to infer or quantify the affairs of
the entire network structure from partial observations. Specifically, many
affiliation relationships of socio-economic systems can be expressed as a
bipartite network. Describing the network structure of complex systems that 
consist of two types of nodes by using the bipartite network
is important. A bipartite graph model also can be used as a general
model for complex networks~\cite{Guillaume:06,Holyst:07,Tumminello}.
Tumminello et al. proposed a statistical method to validate the
heterogeneity of bipartite networks~\cite{Tumminello}. 

Suppose a symmetric binary two-mode network can be
constructed by linking $K$ groups (A node) and $M$ participants (B node) 
if the participants belong to groups. Assume that we can count the number of
participants in each group within the time window $[t\delta,
  (t+1)\delta] \quad (t=1,2,3,\ldots)$, which is defined as $m_i(t)
\quad (i=1,2,\ldots,K)$. 

Let us assume a bipartite graph consisting of $A$ nodes and $B$
nodes, of which the structure at time $t$ is described as an adjacency
matrix $C_{ij}(t)$. We also assume that $A$ nodes are observable and $B$
nodes are unobservable. That is, we only know the number of
participants ($B$ node) belonging to $A$ nodes $m_i(t)$. We do not
know the correct number of $B$ nodes, but we assume that it is
$M$. In this setting, how do we measure the complexity of the bipartite
graph from $m_i(t)$ at each observation time $t$? 

The network entropy is defined as a logarithmic form of the number of
possible configurations of a network under a
constraint~\cite{Bianconi}. We can introduce the network entropy
at time $t$ as a measure to quantify the complexity of a bipartite network
structure. The number of possible configurations under the constraint
$m_i(t)=\sum_{j=1}^MC_{ij}(t)$ may be counted as
\begin{equation}
N(t) = \prod_{i=1}^K
\left(
\begin{array}{c}
M \\
m_i(t)
\end{array}
\right) = \prod_{i=1}^K\frac{M!}{m_i(t)!(M-m_i(t))!}.
\label{eq:NN}
\end{equation}
Then, the network entropy is defined as $\Sigma(t)=\ln N(t)$. Inserting
Eq. (\ref{eq:NN}) into this definition, we have
\begin{equation}
\Sigma(t) = K \sum_{n=1}^M\ln n - \sum_{i=1}^K \sum_{n=1}^{M-m_i(t)} \ln n -
 \sum_{i=1}^K \sum_{n=1}^{m_i(t)} \ln n. 
\label{eq:combination}
\end{equation}
Note that, because $0! = 1$, $\sum_{n=1}^{0} \ln n =0$. Obviously, if
$m_i(t)=M$ for any $i$, then $\Sigma(t) = 0$. If $m_i(t)=0$ for any $i$, then
$\Sigma(t) = 0$. The lower number of combinations gives a lower value of
$\Sigma(t)$. To eliminate a difference in the number of links, we consider
the network entropy per link defined as
\begin{equation}
\sigma[m_1(t),\ldots,m_K(t)] = \frac{\Sigma(t)}{\sum_{i=1}^K m_i(t)}.
\label{eq:sigma}
\end{equation}
This quantity shows the degree of complexity of the bipartite network
structure. We may capture the temporal development of the network
structure from the value of $\sigma(t)$. The network entropy per link
$\sigma(t)$ is also an approximation of the ratio of the entropy rate for
$m_i(t)$ to its mean so that
\begin{equation}
\sigma(t) =
\frac{\frac{1}{K}\Sigma[m_1(t),\ldots,m_K(t)]}{\frac{1}{K}\sum_{i=1}^K
  m_i(t)} \approx \frac{\Sigma[{\bf m}(t)]}{\langle m(t) \rangle},
\end{equation}
where the entropy rate and the mean are, respectively, defined as
\begin{eqnarray}
\Sigma[{\bf m}(t)] &=& \lim_{K\rightarrow
  \infty}\frac{1}{K}\Sigma[m_1(t),\ldots,m_K(t)], \\
\langle m(t) \rangle &=&
\lim_{K\rightarrow\infty}\frac{1}{K}\sum_{i=1}^K m_i(t).
\end{eqnarray}
The ratio of the entropy rate to the mean tells us the uncertainty of
the mean from a different point of view from the coefficient of
variation ($C.V. = \mbox{standard deviation} / \mbox{mean}$).

\begin{figure}[hbt]
\includegraphics[scale=0.62]{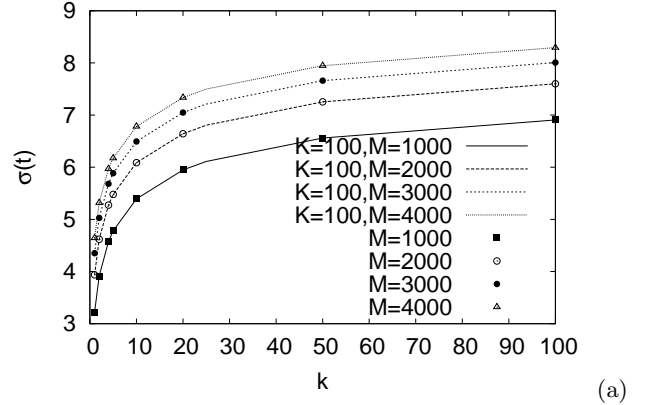}(a)
\includegraphics[scale=0.62]{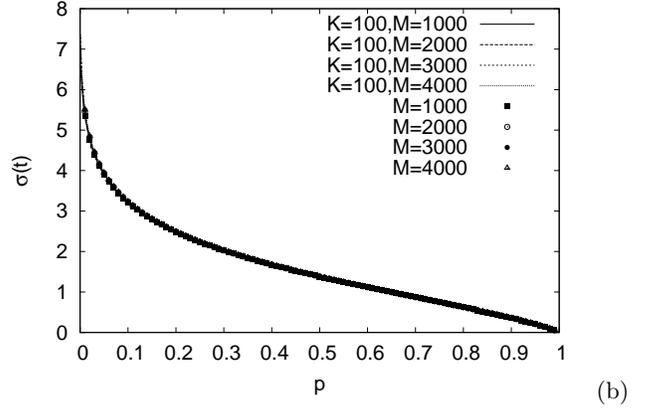}(b)
\caption{(a) Plots between $\sigma(t)$ and degree of monopolization
 $k$. Each curve represents the relation between $\sigma(t)$ and
  $k$. Filled squares numerical values for $M=1000$, unfilled circles
  for $M=2000$, filled circles for $M=3000$, and unfilled triangle for
  $M=4000$. (b) Plots between $\sigma(t)$ and density of links $p$.
  Each curve represents the relation between $\sigma(t)$ and
  $k$. Filled squares numerical values for $M=1000$, unfilled circles
  for $M=2000$, filled circles for $M=3000$, and unfilled triangle for
  $M=4000$.}
\label{fig:denbin}
\end{figure}

To understand the fundamental properties of Eq. (\ref{eq:sigma}), 
we compute $\sigma(t)$ in simple cases. Consider values of
entropy for several cases at $K=100$ with different $M$. We assume that the
total number of links is fixed at 100, which is the same as the number
of $A$ nodes, and we confirm the dependence of $\sigma(t)$ on the degree of
monopolization. We assign the same number of links at each $A$
node. That is, we set 
\begin{equation}
m_i(t) =
\left\{
\begin{array}{ll}
100/k & i=1,\ldots,k \\
0 & i=k+1,\ldots,K
\end{array}
\right.,
\end{equation}
where $k$ can be set as 1, 2, 4, 5, 10, 20, 50, or 100. In this case,
we can calculate $\sigma(t)$ as follows:
\begin{eqnarray}
\nonumber
\sigma(t) &=& \frac{\sum_{i=1}^k \ln\left(\begin{array}{c}M
    \\ 100/k \end{array}\right)}{\sum_{i=1}^k 100/k} \\
&=& \frac{k}{100}\Bigl(\ln M! - \ln (100/k)! - \ln(M-100/k)! \Bigr).
\end{eqnarray}

Fig. \ref{fig:denbin} a shows the relationship between $\sigma(t)$
and the degree of monopolization at $M=1,000$, $2,000$, $3,000$, and
$4,000$. The network entropy per link $\sigma(t)$ is small if a small
population of nodes occupies a large number of links. The multiplication
regime gives a large value of $\sigma(t)$. The value of $\sigma(t)$ is
a monotonically increasing function in terms of $k$. As $M$ increases, the
value of $\sigma(t)$ increases. From this instance, we
confirmed that $\sigma(t)$ decreases with the degree of monopolization at
$A$ nodes. 

Next, we confirm the dependency of $\sigma(t)$ on the density
of links. We assume that each element of an adjacency matrix $C_{ij}(t)$
is given by an {\it i.i.d.} Bernoulli
random variable with a successful probability of $p$. Then,
$m_i(t)=\sum_{j=1}^MC_{ij}(t)$ is sampled from an {\it i.i.d.} 
binomial distribution $Bin(p,M)$. In this case, one can approximate
$\sigma(t)$ as
\begin{eqnarray}
\nonumber
\sigma(t) &=& \frac{\frac{1}{K}\sum_{i=1}^K \Sigma[m_i(t)] }{\frac{1}{K}\sum_{i=1}^K m_i(t)} \\ 
\nonumber
& \approx & \frac{\langle \Sigma[m_1(t)] \rangle}{\langle m_1(t) \rangle} \\
&=& \frac{1}{M}\sum_{k=1}^M \left(\begin{array}{c}M \\ k\end{array}\right)p^{k-1}(1-p)^{M-k}\ln\left(\begin{array}{c}M \\ k\end{array}\right)
\label{eq:binomial-theory}
\end{eqnarray}

Fig. \ref{fig:denbin} b shows the plots of
$\sigma(t)$ versus $p$ obtained from both Monte Carlo simulation with
random links drawn from Bernoulli trials and
Eq. (\ref{eq:binomial-theory}). The number of links at each $A$ node
monotonically increases as $p$ increases. $\sigma(t)$ decreases
as the density of links decreases. The dependence of the entropy per
link on $p$ is independent of $M$.

\section{Empirical analysis}
\label{sec:empirical}
The application of network analysis to financial time series has
been advancing. Several researchers have investigated the network
structure of financial markets~\cite{Bonanno:03,Gworek:09,Podnik:09,Iori:08}. 
Bonanno et al. examined the topological
characterization of the correlation-based 
minimum spanning tree (MST) of real data~\cite{Bonanno:03}. Gworek et
al. analyzed the exchange rate returns of 38 currencies (including
gold) and computed the characteristic path length and average weighted
clustering coefficient of the MST topology of the graph extracted from
the cross-correlations for several base
currencies~\cite{Gworek:09}. Podnik et al.~\cite{Podnik:09} examined the
cross-correlations between volume changes and price changes for the New
York Stock Exchange, Standard and Poor's 500 index, and 28 worldwide
financial indices. Iori et al.~\cite{Iori:08} analyzed the network
topology of the Italian segment of the European overnight money market
and investigated the evolution of these banks' connectivity structure
over the maintenance period. These studies collectively aimed to detect the
susceptibility of network structures to macroeconomic situations.

Data collected from the ICAP EBS platform were used. The data
period spanned May 28, 2007 to November 30, 2012~\cite{ICAP}. 
The data included records for orders
(BID/OFFER) and transactions for currencies and precious metals with
a one-second resolution. The data set involved 94 currency pairs
consisting of 39 currencies, 11 precious metals, and 2 basket currencies 
(AUD, NZD, USD, CHF, JPY, EUR, CZK, DKK, GBP, HUF, ISK, NOK, PLN, SEK,
SKK, ZAR, CAD, HKD, MXC, MXN, MXT, RUB, SGD, XAG, XAU, XPD, XPT, TRY,
THB, RON, BKT, ILS, SAU, DLR ,KES, KET, AED, BHD, KWD, SAR, EUQ, USQ,
CNH, AUQ, GBQ, KZA, KZT, BAG, BAU, BKQ, LPD, and LPT) 

\subsection{The total number}
\begin{figure}[hbt]
\includegraphics[scale=0.6]{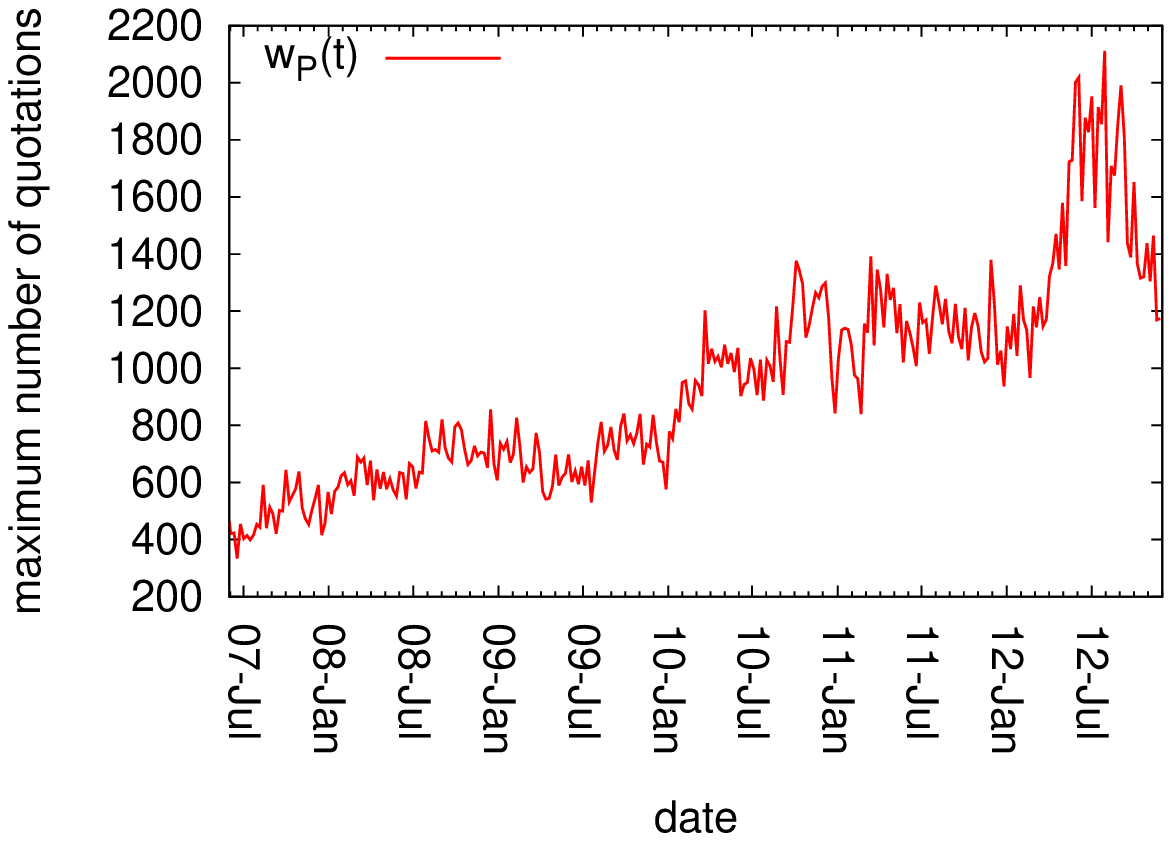}(a)
\includegraphics[scale=0.6]{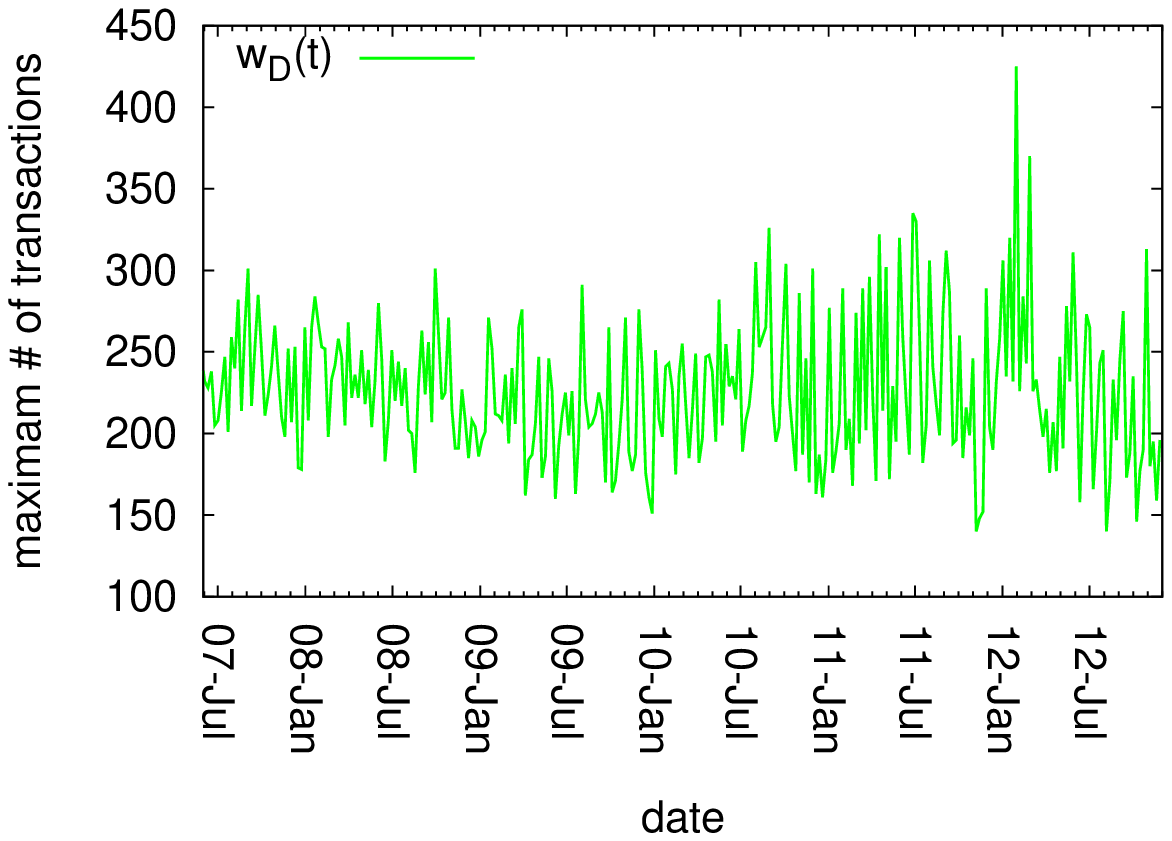}(b)
\caption{(a) The maximum values of the number of quotations within
  1 minute in every week. (b) the maximum values of the number of
transactions within 1 minute in every week.}
\label{fig:maxcount}
\end{figure}

The number of quotations and transactions in each currency pair was
extracted from the raw data. Let $m_{X,i}(t) \quad
(t=0,\ldots;i=1,\ldots,K)$ be the number of quotations ($X=P$) or
transactions ($X=D$) within every minute ($\delta = 1$ [min]) for 
a currency pair $i$ ($K=94$) at time $t$. Let $c_{X}(t)$ be denoted 
as the total number of quotations ($X=P$) and transactions ($X=D$),
which is defined as
\begin{equation}
c_{X}(t) = \sum_{i=1}^K m_{X,i}(t).
\end{equation}
Let us consider the maximum value of $c_X(t)$ in each week: 
\begin{equation} 
w_X(s) = \max_{t \in W(s)}\{ c_X(t) \},
\end{equation}
where $W(s) \quad (s=1, \ldots, T)$ represents a set of times included in
the $s$-th week. A total of 288 weeks are included in the data set
($T=288$). Fig. \ref{fig:maxcount} shows the maximum values $c_X(t)$
for the period from May 28, 2007 to November 30, 2012.

According to the extreme value theorem, the probability density
for maximum values can be assumed to be a Gumbel density: 
\begin{equation}
P(w_X; \mu_X, \rho_X) =
 \frac{1}{\rho_X}\exp\Bigl(-\frac{w_X-\mu_X}{\rho_X}-e^{-\frac{w_X-\mu_X}{\rho_X}}\Bigr), 
\label{eq:Gumbel-max}
\end{equation}
where $\mu_X$ and $\rho_X$ are the location and scale parameters,
respectively. Under the assumption of the Gumbel density, these parameters are
estimated with the maximum likelihood procedure. The log--likelihood
function for $T$ observations $w_X(s') \quad (s'=1,\ldots,T)$ under
Eq. (\ref{eq:Gumbel-max}) is defined as
\begin{equation}
l(\mu_X,\rho_X) = \sum_{s'=1}^T \ln\Bigl[
\frac{1}{\rho_X}\exp\Bigl(-\frac{w_X(s')-\mu_X}{\rho_X}-e^{-\frac{w_X(s')-\mu_X}{\rho_X}}\Bigr)
\Bigr]. 
\end{equation}
The maximum likelihood estimators are obtained by maximizing
the log-likelihood function. Partially differentiating
$l(\mu_X,\rho_X)$ in terms of $\mu_X$ and $\rho_X$ and setting them
to zero, one has its maximum likelihood estimators as
\begin{eqnarray}
e^{-\frac{\hat{\mu}_X}{\hat{\rho}_X}} &=& \frac{1}{T} \sum_{s'=1}^T
e^{-\frac{w_X(s')}{\hat{\rho}_X}},
\label{eq:MLE-mu-max}
\\
\hat{\rho}_X &=& \frac{1}{T}\sum_{s'=1}^T w_X(s') -
\frac{\sum_{s'=1}^T
  e^{-\frac{w_X(s')}{\hat{\rho}_X}}w_X(s')}{\sum_{s'=1}^T e^{-\frac{w_X(s')}{\hat{\rho}_X}}} 
\label{eq:MLE-rho-max}
\end{eqnarray}
Their derivation is shown in Appendix \ref{sec:derivation-max}. The
parameters are estimated as $\hat{\mu}_P = 772.179499$, $\hat{\rho}_P =
281.741815$, $\hat{\mu}_D = 206.454884$, and $\hat{\rho}_D = 35.984804$.

The Kolmogorov--Smirnov (KS) test is conducted to determine the
statistical significance of the estimated distributions. The KS test
is a popular statistical method of assessing the difference between
observations and its assumed distribution by p-value, which is a measure
of probability where a difference between the two distributions
happens by chance. Large p-values imply that the observations are
sampled from the assumed distribution in the null hypothesis with high
significance. Let $w_X(s) \quad (s=1,\ldots,T)$ be $T$ observations, and
let $K_{T}$ be a test statistic 
\begin{equation}
\nonumber
K_{T} = \sup_{1 \le s' \le T}\sqrt{T}\Bigl|F_{T}(w_{X}(s'))-F(w_{X}(s'))\Bigr|, 
\label{eq:distance}
\end{equation}
where $0 \leq F(v) \leq 1$ is an assumed cumulative distribution in a
null hypothesis and $F_{T}(v)$ an empirical one based on $T$
observations such that $F_{T}(v)=k/T$, in which $k$
represents the number of observations satisfying 
$v_{X}(s) \leq v (s=1,\ldots,T)$. The p-value is computed from
the Kolmogorov--Smirnov distribution.

The KS test is conducted under the assumption of the
Gumbel distribution for the maximum value corresponding to
Eq. (\ref{eq:Gumbel-min}):
\begin{equation}
F(w_X; \mu_X, \rho_X) = \exp\Bigl[-\exp\bigl(-\frac{w_X - \mu_X}{\rho_X}\bigr)\Bigr].
\end{equation}
The p-values of the KS test are shown in Tab.
\ref{tab:p-value-for-stationary-max}. The stationary Gumbel assumption
cannot explain the maximum values for quotes with a 5\% significance level
in the KS test. The stationary Gumbel assumption may not be accepted
in the case of the block maximum number of quotes. The dominant
reason is the strong nonstationarity of the maximum number of
quotes. During the last five years, the currencies and pairs
quoted in the electronic brokerage market increased. The mean value of
the total number constantly increased. In fact, the maximum number of
quotations $w_P(t)$ reached the maximum value on 30 July, 2012. The
nonstationarity breaks the assumption of the extreme value theorem. 

It is confirmed that the stationary Gumbel assumption can be accepted
for the block maxima of transactions in each week using the KS test
with a 5\% significance level. The maximum number of transactions
$w_D(t)$ was reached on on January 30, 2012. This period seems
to be related to the extreme synchrony. 

\begin{table}[hbt]
\caption{The p-values of statistical tests under a stationary 
assumption of the Gumbel distribution for the maximum values.}
\label{tab:p-value-for-stationary-max}
\begin{tabular}{llll}
\hline
p-val (P) & KS-val (P) & p-val (D) & KS-val (D) \\
\hline
0.041374 & 1.392521 & 0.586818 & 0.774087 \\
\hline
\end{tabular}
\end{table}

\subsection{Network entropy per link}
The proposed method based on statistical--physical entropy is applied
to measure the states of the foreign exchange market. The relationship
between a bipartite network structure and macroeconomic shocks or
crises was investigated, and the occurrence probabilities of extreme 
synchrony were inferred. We compute a statistical--physical entropy
per link from $m_{X,i}(t) (X \in \{P,D\})$ with
Eqs. (\ref{eq:combination}) and (\ref{eq:sigma}), which are denoted as
$\sigma_X(t)$. $\sigma_P(t)$ and $\sigma_D(t)$.

Since small values of $\sigma_X(t)$ correspond to a concentration of
links at a few nodes or a dense network structure, let us consider the
minimum value of $\sigma_X(t)$ every week:
\begin{equation} 
v_X(s) = \min_{t \in W(s)}\{\sigma_X(t)\},
\end{equation}
where $W(s) \quad (s=1, \ldots, T)$ represents a set of times included in
the $s$-th week. A total of 288 weeks are included in the data set
($T=288$). According to the extreme value theorem, the probability density
for minimum values can be assumed to be the Gumbel density: 
\begin{equation}
P(v_X; \mu_X, \rho_X) =
 \frac{1}{\rho_X}\exp\Bigl(\frac{v_X+\mu_X}{\rho_X}-e^{\frac{v_X+\mu_X}{\rho_X}}\Bigr), 
\label{eq:Gumbel-min}
\end{equation}
where $\mu_X$ and $\rho_X$ are the location and scale parameters,
respectively. Under the assumption of the Gumbel density, these parameters are
estimated with the maximum likelihood procedure. The log--likelihood
function for $T$ observations $v_X(s') \quad (s'=1,\ldots,T)$ under
Eq. (\ref{eq:Gumbel-min}) is defined as
\begin{equation}
l(\mu_X,\rho_X) = \sum_{s'=1}^T \ln\Bigl[
\frac{1}{\rho_X}\exp\Bigl(\frac{v_X(s')+\mu_X}{\rho_X}-e^{\frac{v_X(s')+\mu_X}{\rho_X}}\Bigr)
\Bigr]. 
\end{equation}
Partially differentiating $l(\mu_X,\rho_X)$ in terms of $\mu_X$ and $\rho_X$
and setting them to zero yields its maximum likelihood estimators as
\begin{eqnarray}
e^{-\frac{\hat{\mu}_X}{\hat{\rho}_X}} &=& \frac{1}{T} \sum_{s'=1}^T
e^{\frac{v_X(s')}{\hat{\rho}_X}},
\label{eq:MLE-mu-min}
\\
\hat{\rho}_X &=& \frac{\sum_{s'=1}^T e^{\frac{v_X(s')}{\hat{\rho}_X}}v_X(s')}{\sum_{s'=1}^T
  e^{\frac{v_X(s')}{\hat{\rho}_X}}} - \frac{1}{T}\sum_{s'=1}^T v_X(s').
\label{eq:MLE-rho-min}
\end{eqnarray}
These derivations are shown in Appendix \ref{sec:derivation-min}.
The parameter estimates are computed as $\hat{\mu}_P=-4.865382$, 
$\hat{\rho}_P=0.110136$, $\hat{\mu}_D=-5.010175$, and
$\hat{\rho}_D=0.120809$ with Eqs. (\ref{eq:MLE-mu-min}) and
(\ref{eq:MLE-rho-min}). 

The KS test is conducted for the Gumbel distribution for
the minimum values corresponding to Eq. (\ref{eq:Gumbel-min}):
\begin{equation}
F(v_X; \mu_X, \rho_X) = 1 - \exp\Bigl[-\exp\bigl(\frac{v_X+\mu_X}{\rho_X}\bigr)\Bigr].
\end{equation}
The p-value of the distribution is shown in Tab. 
\ref{tab:p-value-for-stationary-min}. The stationary Gumbel
assumption cannot explain the synchronizations observed in both
quotes and transactions completely with a 5\% significance
level. The stationary Gumbel assumption is
rejected because there is a stationary assumption to derive the
extreme value distribution. If we can weaken this assumption, then the
goodness of fit may be improved.
\begin{table}[hbt]
\caption{The p-values of statistical tests under a stationary Gumbel
  assumption.}
\label{tab:p-value-for-stationary-min}
\begin{tabular}{llll}
\hline
p-val (P) & KS-val (P) & p-val (D) & KS-val (D) \\
\hline
0.001393 & 1.906528 & 0.019241 & 1.523791 \\
\hline
\end{tabular}
\end{table}

\section{Probability of extreme synchrony}
\label{sec:probability}
The literature detecting structural breaks or change points in an economic
time series~\cite{Goldfeld,Tobias,Scalas,Cheong} points out that
nonstationary time series are constructed from locally stationary
segments sampled from different distributions. 
Goldfeld and Quandt conducted a pioneering work on the separation
of stationary segments~\cite{Goldfeld}. Recently, a hierarchical
segmentation procedure was also proposed by Choeng et al.
under the Gaussian assumption~\cite{Cheong}. We applied this concept to
define the segments for $v_X(s') \quad (s'=1,\ldots,T)$. 

Let us consider the null model $L_1$, which assumes that all the
observations $v_X(s') \quad (s'=1,\ldots,T)$ are sampled from a
stationary Gumbel density parameterized as $\mu$ and $\rho$.
An alternative model $L_2(s)$ assumes that the left observations
$v_X(s') \quad (s'=1,\ldots,s)$ are sampled from a
stationary Gumbel density parameterized as $\mu_L$ and $\rho_L$, and that 
the right observations $v_X(s') \quad (s'=s+1,\ldots,T)$ are sampled
from a stationary Gumbel density parameterized as $\mu_R$ and $\rho_R$.

Denoting likelihood functions as
\begin{eqnarray}
L_1(\mu,\rho) &=& \prod_{s'=1}^T P(v_X(s');\mu,\rho), \\
\nonumber
L_2(s;\mu_L,\rho_L,\mu_R,\rho_R) &=& \prod_{s'=1}^s
P(v_X(s');\mu_{L},\rho_{L}) \\
&\times& \prod_{s'=s+1}^{T} P(v_X(s');\mu_{R},\rho_{R}),
\end{eqnarray}
the difference between the log--likelihood functions can be defined as
\begin{equation}
\Delta(s) = \log L_2(s) - \log L_1.
\end{equation}
$\Delta(s)$ can be approximated as the Shannon entropy
$H[p]=-\int_{\infty}^{\infty}\mbox{d}v \log p(v)p(v)$: 
\begin{eqnarray}
\nonumber
\Delta(s) &\approx& T H[P(v_X;\mu,\rho)] \\
\nonumber
&-& s H[P(v_X;\mu_L,\rho_L)] - (T-s) H[P(v_X;\mu_R,\rho_R)].\\
\end{eqnarray}
Since the Shannon entropy of the Gumbel density expressed in
Eq. (\ref{eq:Gumbel-max}) is calculated as 
\begin{equation}
H[P(v_X;\mu_X,\rho_X)] = \ln \rho_X - \gamma + 1,
\end{equation}
where $\gamma$ represents Euler's constant, defined as
\begin{equation}
\gamma = \int_{0}^{\infty} \ln t e^{-t}\mbox{d}t,
\end{equation}
we obtain
\begin{equation}
\Delta(s) \approx T\ln \rho - s\ln \rho_L - (T-s)\ln \rho_R.
\end{equation}

In the context of model selection, several information criteria are
proposed. The information criterion provides both goodness of fit of
the model to the data and model complexity. For the sake of simplicity,
we use the Akaike information criterion (AIC) to determine the adequate
model. The AIC for a model with the number of parameters $K$ and the
maximum likelihood of $L$ is defined as
\begin{equation}
AIC = -2 \ln L + 2K.
\end{equation}
We can compute the difference in AIC 
between model $L_2$ and model $L_1(s)$ as 
\begin{eqnarray}
\nonumber
\Delta_{AIC}(s) &=& \mbox{AIC of $L_2(s)$} - \mbox{AIC of $L_1$} \\
\nonumber
&\approx& -2\bigl(T\ln \hat{\rho} - s\ln \hat{\rho}_L - (T-s)\ln
\hat{\rho}_R\bigr) + 4, \\
&=& -2\Delta(s) + 4
\end{eqnarray}
since the number of parameters of $L_1$ is 2, that of $L_2(s)$ is 4,
and the maximum likelihood is obtained by using their maximum
likelihood estimators calculated from 
\begin{eqnarray}
\hat{\rho} &=& \frac{\sum_{s'=1}^T e^{\frac{v_X(s')}{\hat{\rho}_X}}v_X(s')}{\sum_{s'=1}^T
  e^{\frac{v_X(s')}{\hat{\rho}_X}}} - \frac{1}{T}\sum_{s'=1}^Tv_X(s') \\
\hat{\rho}_L &=& \frac{\sum_{s'=1}^s e^{\frac{v_X(s')}{\hat{\rho}_L}}v_X(s')}{\sum_{s'=1}^s
  e^{\frac{v_X(s')}{\hat{\rho}_L}}} - \frac{1}{s}\sum_{s'=1}^{s}v_X(s') \\
\nonumber
\hat{\rho}_R &=& \frac{\sum_{s'=s+1}^T
  e^{\frac{v_X(s')}{\hat{\rho}_R}}v_X(s')}{\sum_{s'=s+1}^T 
  e^{\frac{v_X(s')}{\hat{\rho}_R}}} - \frac{1}{T-s}\sum_{s'=s+1}^Tv_X(s') \\
\end{eqnarray}

Therefore, $P(v_X;\mu_{L},\rho_{L})$ is maximally different from
$P(v_X;\mu_{R},\rho_{R})$ when $\Delta(s)$ assumes a maximal
value. This spectrum has a maximum at some time $s^{*}$, which
is denoted as
\begin{equation}
\Delta^*_{AIC} = \Delta_{AIC}(s^*) = \max_{s} \Delta_{AIC}(s).
\end{equation}
The segmentation can be used recursively to separate the time series into
further smaller segments. We do this iteratively until all segment
boundaries have converged onto their optimal segment, defined by a
stopping (termination) condition. 

Several termination conditions were discussed in previous 
studies~\cite{Cheong}. Assuming that $\Delta_0 > 0$, we terminate the
iteration if $\Delta_{AIC}^{*}$ is less than a typical conservative
threshold of $\Delta_0=10$, while the procedure is recursively
conducted if $\Delta_{AIC}^*$ is larger than $\Delta_0$. We checked
the robustness of this segmentation procedure for
$\Delta_0$. $\Delta_0$ gives a statistical significance level of
termination. The value of $\Delta_0$ is related to statistical
significance. According to Wilks theorem, $-2\Delta(s)$ follows
a chi-squared distribution with a degree of freedom $r$, where $r$ is
given by the difference between the number of parameters assumed in
the null hypothesis and one in the alternative hypothesis. In this
case, $r=2$. Hence, the cumulative distribution function of
$\Delta^*_{AIC}$ may follow
\begin{equation}
\mbox{Pr}[\Delta^*_{AIC} > x] = 1- \gamma\Bigl(1,\frac{x-4}{2}\Bigr).
\end{equation}
Therefore, setting the threshold $\Delta_0=10$ implies that the
segmentation procedure is tuned as a 4.928\% statistical significance
level.

Let the number of segments be $L_X$, the parameter estimates be $\{\mu_{X,j},
\rho_{X,j}\}$ at the $j$-th segment, and the length of the $j$-th
segment be $\tau_{X,j}$, where $\sum_{j=1}^{R_X}\tau_{X,j}=T$. The cumulative
probability distribution for $v_X(s) \quad (s=1,\ldots,T)$ may be
assumed to be a finite mixture of Gumbel distributions:
\begin{eqnarray}
\nonumber
\mbox{Pr}(V_X\leq v_X) &=& \int_{-\infty}^{v_X}\sum_{j=1}^{R_X}\frac{\tau_{X,j}}{T}P(v_X';\mu_{X,j},\rho_{X,j})\mbox{d}v_X' \\
&=& \sum_{j=1}^{R_X} \frac{\tau_{X,j}}{T}
 \Bigl\{1-\exp\bigl[-e^{\frac{v_X+\mu_{X,j}}{\rho_{X,j}}}\bigr]\Bigr\},
\label{eq:mixture-CDF}
\end{eqnarray}

\begin{figure}[hbt]
\centering
\includegraphics[scale=0.66]{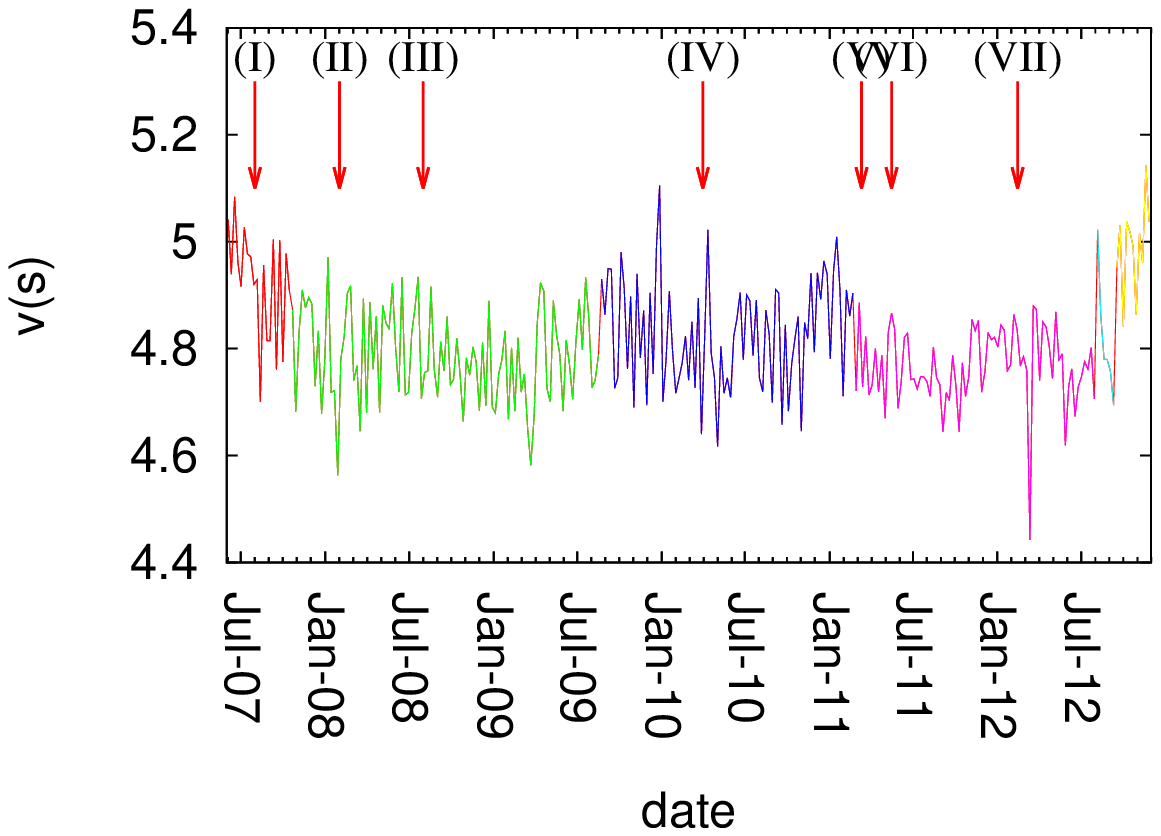}(a)
\includegraphics[scale=0.66]{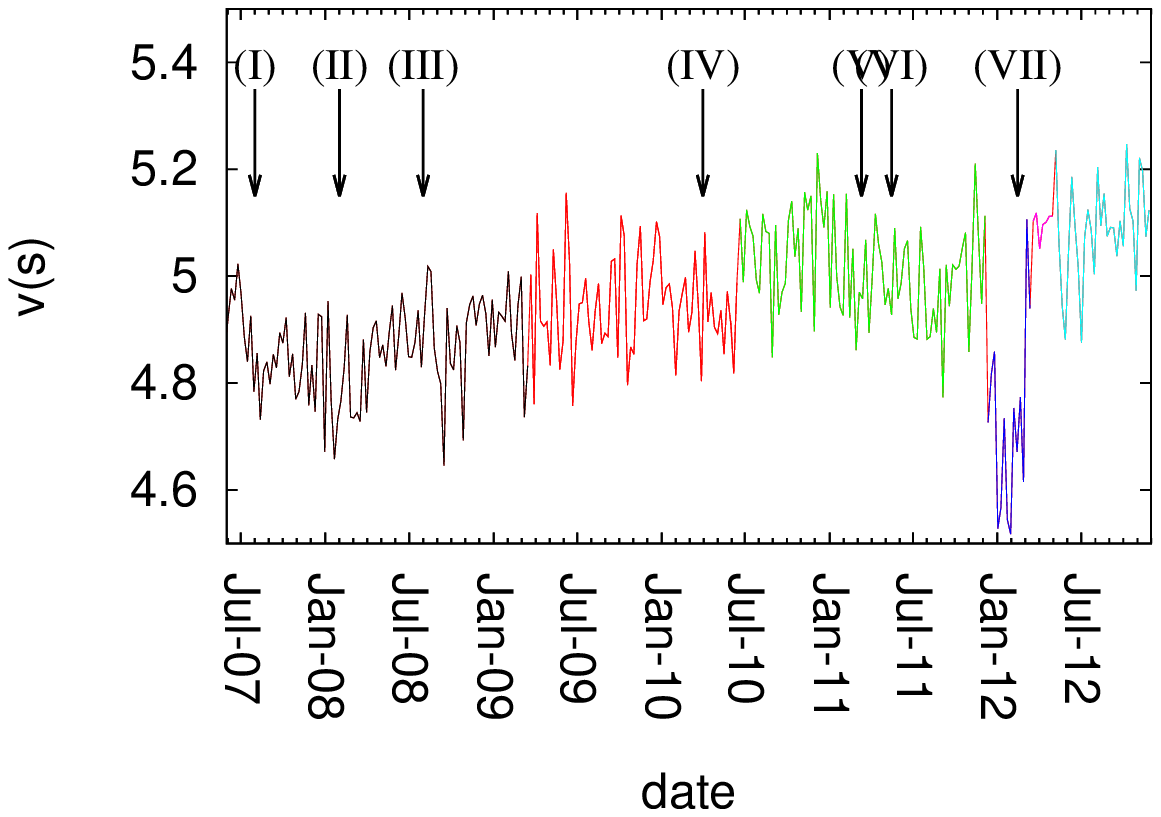}(b)
\caption{Temporal development of (a) $v_P(s)$ and (b) $v_D(s)$
from May 28, 2007 to November 30, 2012.}
\label{fig:minpd}
\end{figure}

Tabs. \ref{tab:minp} and \ref{tab:mind} show parameter estimates of
$v_P(s)$ and $v_D(s)$ using the recursive segmentation procedure.
\begin{table}[hbt]
\caption{Parameter estimates obtained from the weekly minimum values of
  network entropy for quotations with the recursive segmentation
  procedure.}
\label{tab:minp}
\begin{tabular}{llllll}
\hline
$j$ & start date & end date & $\tau/T$ & $\hat{\mu}_{X,j}$ & $\hat{\rho}_{X,j}$ \\
\hline
1 & May 28, 2007 & Oct. 15, 2007 & 0.072917 & -4.965486 & 0.078571 \\
2 & Oct. 22, 2007 & Aug. 17, 2009 & 0.333333 & -4.829061 & 0.080558 \\
3 & Aug. 24, 2009 & Oct. 21, 2011 & 0.274306 & -4.880156 & 0.099532 \\
4 & Feb. 28, 2011 & Jul. 30, 2012 & 0.260417 & -4.798364 & 0.058265 \\
5 & Aug. 6, 2012 & Sep. 10, 2012 & 0.020833 & -4.856415 & 0.121566 \\
6 & Sep. 17, 2012 & Nov. 26, 2012 & 0.038194 & -5.031930 & 0.071871 \\
\hline
\end{tabular}
\caption{Parameter estimates obtained from the weekly minimum values
  of network entropy for transactions with the recursive segmentation procedure.}
\label{tab:mind}
\begin{tabular}{llllll}
\hline
$j$ & start date & end date & $\tau/T$ & $\hat{\mu}_{X,j}$ & $\hat{\rho}_{X,j}$ \\
\hline
1 & May 28, 2007 & Mar. 16, 2009 & 0.329861 & -4.903660 & 0.072684 \\
2 & Mar. 23, 2009 & Jun. 14, 2010 & 0.225694 & -4.990569 & 0.088420 \\
3 & Jun. 21, 2010 & Dec. 5, 2011 & 0.267361 & -5.065305 & 0.087721 \\
4 & Dec. 12, 2011 & Mar. 12, 2012 & 0.048611 & -4.826384 & 0.174328 \\
5 & Mar. 19, 2012 & Apr. 30, 2012 & 0.024306 & -5.108679 & 0.011079 \\
6 & May 7, 2012 & Nov. 26, 2012 & 0.104167 & -5.131156 & 0.079630 \\
\hline
\end{tabular}
\end{table}
Fig. \ref{fig:minpd} shows the temporal development of $v_P(s)$ and
$v_D(s)$ from May 28, 2007 to November 30, 2012. $R_P=6$ and $R_D=6$ are 
obtained from $v_X(s)$ using the proposed segmentation
procedure. During the observation period, the global financial system
suffered from the following significant macroeconomic shocks and crises:
(I) the BNP Paribas shock (August 2007), (II) the Bear Stearns shock
(February 2008), (III) the Lehman shock (September 2008 to March 2009), 
(IV) the European sovereign debt crisis (April to May 2010), 
(V) the East Japan tsunami (March 2011), (VI) the United States
debt-ceiling crisis (May 2011), and (VII) the Bank of Japan's 10 trillion
JPY gift on Valentine's Day (February 2012).

Before entering these global affairs, both $v_P(s)$ and $v_D(s)$ took large
values. Note that, during the (I) Paribas shock, the (II) Bear
Stearns and the (III) Lehman shock $v_P(s)$ and $v_D(s)$ took smaller
values than they did during the previous term. This implies that
a global shock may drive many participants and that these participants may
trade the same currencies at the same time. The smallest values
$v_P(s)$ and $v_D(t)$ correspond to the days of the (II) Bear Stearns
shock, the (III) Lehman shock, and the (VI) Euro crisis.  
These days are generally related to the start or the end of macroeconomic
shocks or crises. The period from December 2011 to March 2012 shows
that the values of $v_D(s)$ are smaller than they were during other
periods. This result implies that, during said period, singular patterns
appeared in the transactions.

\begin{figure}[!hbt]
\centering
\includegraphics[scale=0.62]{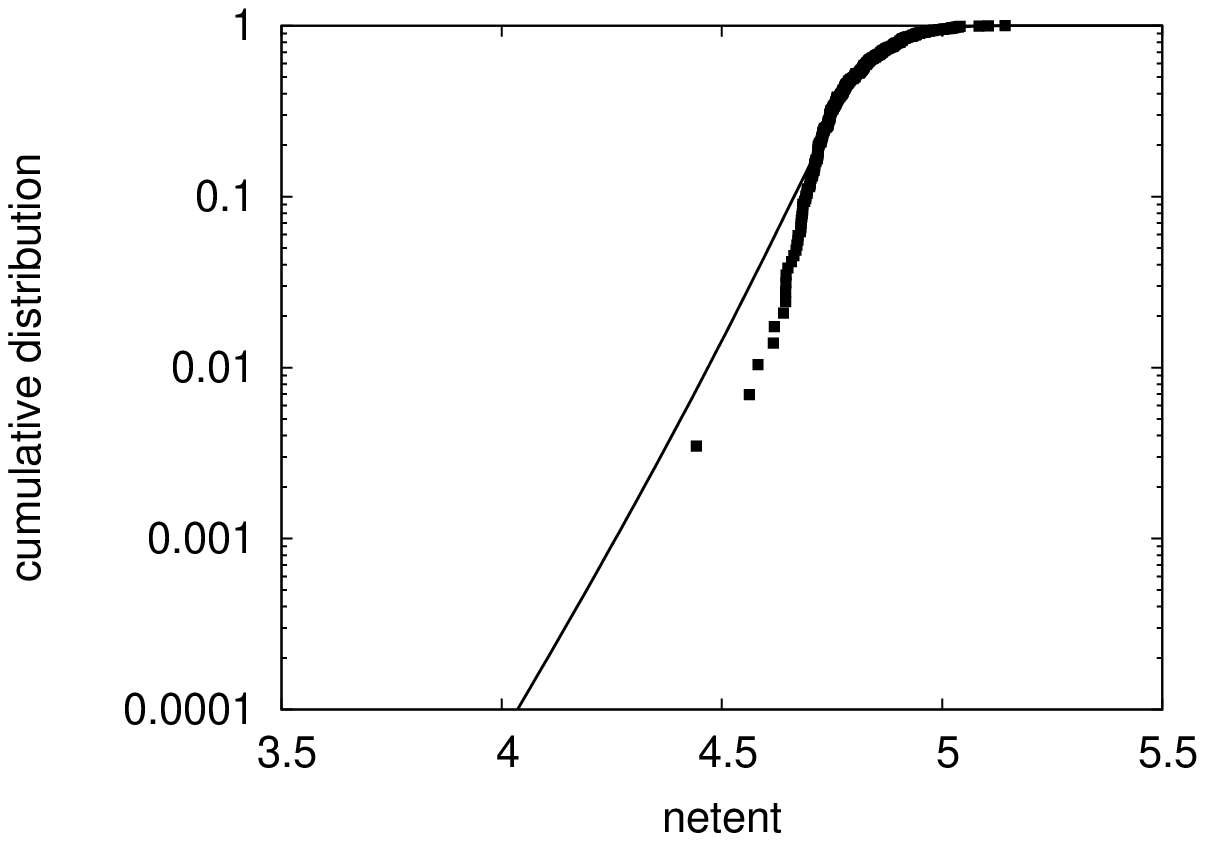}(a)
\includegraphics[scale=0.62]{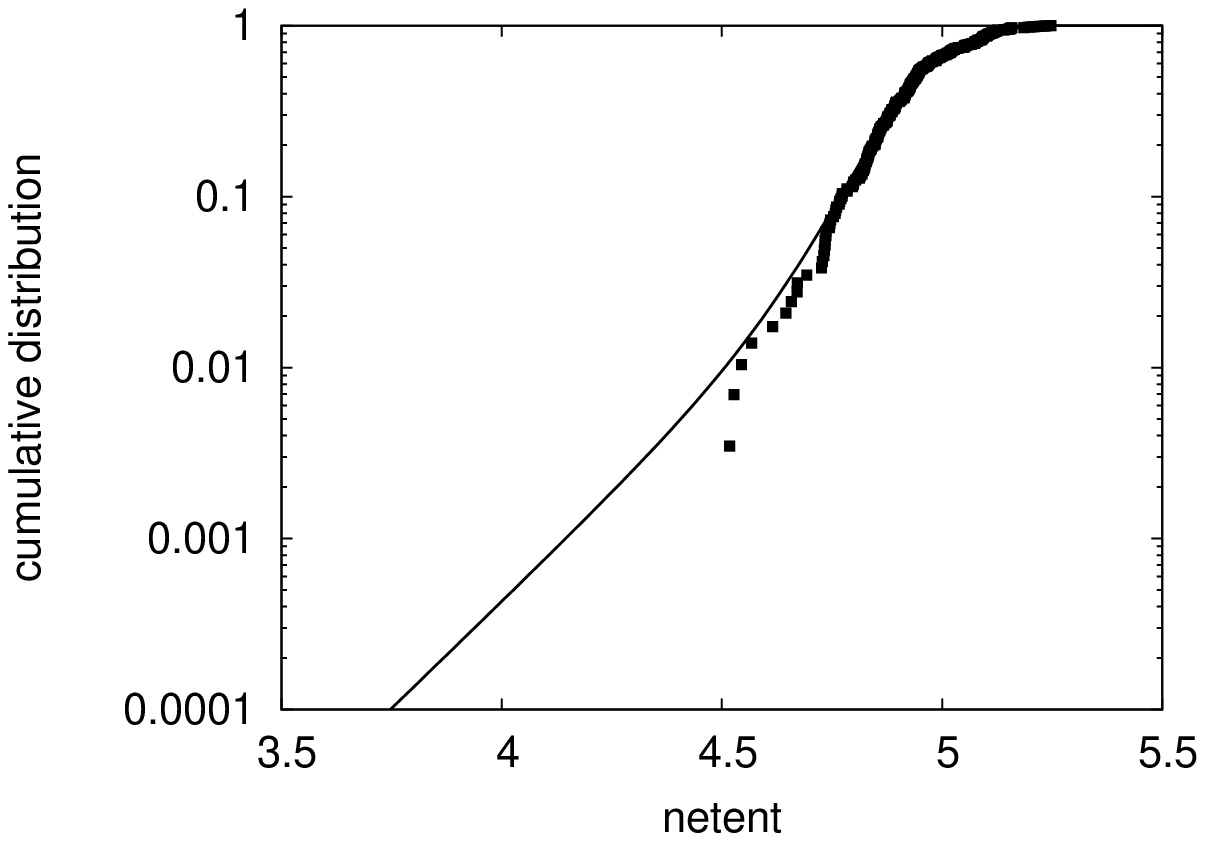}(b)
\caption{Cumulative distribution functions for the minimum values of
 the entropy per link in each week (a) $v_P(s)$ and (b)
 $v_D(s)$. Filled squares represent the empirical distribution of
 $v_P(s)$, and unfilled circles represent the empirical distribution
 of $v_D(s)$. A solid curve represents the estimated distribution of
 $v_P(s)$, and a dashed curve represents the estimated distribution of
 $v_D(s)$.}
\label{fig:mincdf}
\end{figure}

Fig. \ref{fig:mincdf} shows both the empirical and estimated
cumulative distribution functions of $v_P(s)$ and
$v_D(s)$. The estimated cumulative distributions are drawn from
Eq. (\ref{eq:mixture-CDF}) with parameter
estimates. The KS test verifies this mixing
assumption. The distribution estimated by the finite mixture of Gumbel 
distributions for quotes is well fitted, as shown in
Tab. \ref{tab:p-value-for-mixture}. From the $p$-values, the mixture
of Gumbel distributions for quotations accepts the null hypothesis that
$v_P(s)$ is sampled from the mixing distribution with a 5\%
significance level. The mixture of Gumbel distributions for
transactions also accepts the null hypothesis that $v_D(s)$ is sampled
from the mixing distribution with a 5\% significance
level. Extrapolation of cumulative distribution function also provides a
guideline of the future probability of extreme events.
The finite mixture Gumbel distributions with parameter
estimates may be used as an inference of probable extreme synchrony.

\begin{table}[hbt]
\caption{The p-values of statistical tests under the assumption of 
a finite mixture of Gumbel distributions.}
\label{tab:p-value-for-mixture}
\begin{tabular}{llll}
\hline
p-val (P) & KS-val (P) & p-val (D) & KS-val (D) \\
\hline
0.183793 & 1.092317 & 0.829013 & 0.625372 \\
\hline
\end{tabular}
\end{table}

\section{Conclusion}
\label{sec:conclusion}
A method based on the concept of ``entropy'' in statistical
physics was proposed to quantify states of a bipartite network
under constraints. The statistical--physical network entropy 
of a bipartite network was derived under the constraints for the number
of links at each group node. Both numerical and theoretical
calculations for a binary bipartite graph with random links
showed that the network entropy per link can
capture both the density and the concentration of links in the
bipartite network. The proposed method was applied to measure
the structure of bipartite networks consisting of currency pairs and
participants in the foreign exchange market. 

An empirical investigation of the total number of quotes
and transactions was conducted. The nonstationarity of the number of
quotes and transactions strongly affected the extreme value
distributions. The empirical investigation confirmed that the
entropy per link decreased before and after the latest global shocks
that have influenced the world economy. A method was proposed
to determine segments with recursive segmentation based on
the Akaike information criterion between Gumbel distributions with
different parameters. Under the assumption of a finite mixture of
Gumbel distributions, the estimated distributions were verified by 
the Kolmogorov--Smironov test. The finite mixture of Gumbel distributions
can estimate the occurrence probabilities of extreme synchrony of a
nonstationary system extracted as a bipartite network. The
extrapolation of the extreme synchrony can be done based on the
estimated mixture of Gumbel distributions.

\section*{Acknowledgements}
This work was supported by the Grant-in-Aid for Young Scientists (B)
(\#23760074) by the Japanese Society for Promotion of Science (JSPS).
The author expresses his sincere gratitude to Mr. Takashi Isogai 
(Bank of Japan) for his constructive comments.

\appendix
\section{Derivation of the maximum likelihood estimator for Gumbel
  density for the maximum values}
\label{sec:derivation-max}
Take the log--likelihood function of the Gumbel density given by
Eq. (\ref{eq:Gumbel-max}) for $T$ observations $w_X(s') \quad
(s'=1,\ldots,T)$:
\begin{eqnarray}
\nonumber
l(\mu_X,\rho_X) &=& \sum_{s'=1}^T \ln\Bigl[
\frac{1}{\rho_X}\exp\Bigl(-\frac{v_X(s')-\mu_X}{\rho_X}
-e^{-\frac{v_X(s')-\mu_X}{\rho_X}}\Bigr) \Bigr] \\
\nonumber
&=& -T \ln \rho_X - \frac{1}{\rho_X}\sum_{s'=1}^T w_X(s') \\
&+& \frac{\mu_X}{\rho_X}T - \sum_{s'=1}^Te^{-\frac{v_X(s')-\mu_X}{\rho_X}}.
\label{eq:ll-max}
\end{eqnarray}
Partially differentiating Eq. (\ref{eq:ll-max}) in terms of $\mu_X$ and
setting it to zero yields
\begin{eqnarray}
\nonumber
\frac{\partial l}{\partial \mu_X} &=& \frac{T}{\rho_X} -
\sum_{s'=1}^T\frac{1}{\rho_X} e^{-\frac{w_X(s')-\mu_X}{\rho_X}} = 0,\\
\nonumber
T &=& \sum_{s'=1}^T e^{-\frac{w_X(s')-\mu_X}{\rho_X}} \\
e^{-\frac{\mu_X}{\rho_X}} &=& \frac{1}{T} \sum_{s'=1}^T
e^{-\frac{w_X(s')}{\rho_X}}.
\label{eq:eq-of-mu-max}
\end{eqnarray}
Similarly, differentiating Eq. (\ref{eq:ll-max}) in terms
of $\rho_X$ and setting it into zero yields
\begin{eqnarray}
\nonumber
\frac{\partial l}{\partial \rho_X} &=& -\frac{T}{\rho_X} + \frac{1}{\rho_X^2}
\sum_{s'=1}^T v_X(s') - \frac{\mu_X}{\rho_X^2}T \\
\nonumber
&-& \sum_{s'=1}^T
\Bigl(\frac{w_X(s')-\mu_X}{\rho_X^2}e^{-\frac{w_X(s')-\mu_X}{\rho_X}}\Bigr)
= 0, \\ 
\nonumber
-T\rho_X &+& \sum_{s'=1}^T w_X(s') - \mu_X T \\
\nonumber
&-& \sum_{s'=1}^T (w_X(s')-\mu_X) e^{-\frac{w_X(s')}{\rho_X}} = 0, \\
\nonumber
\rho_X + \mu_X &=& \frac{1}{T}\sum_{s'=1}^T w_X(s') -
    e^{-\frac{\mu_X}{\rho_X}} \frac{1}{T} \sum_{s'=1}^T w_X(s')
    e^{-\frac{w_X(s')}{\rho_X}} \\
&-& e^{-\frac{\mu_X}{\rho_X}}
    \frac{\mu_X}{T} \sum_{s'=1}^T e^{-\frac{w_X(s')}{\rho_X}}.
\label{eq:eq-of-rho-max}
\end{eqnarray}
Inserting Eq. (\ref{eq:eq-of-mu-max}) into
Eq. (\ref{eq:eq-of-rho-max}) yields
\begin{eqnarray}
\nonumber
\rho_X + \mu_X &=& \frac{1}{T}\sum_{s'=1}^T w_X(s')
- \frac{\sum_{s'=1}^T w_X(s')
e^{-\frac{w_X(s')}{\rho_X}}}{\sum_{s'=1}^T w_X(s')} + \mu_X \\
\rho_X &=& \frac{1}{T} \sum_{s'=1}^T w_X(s') -
\frac{\sum_{s'=1}^T e^{-\frac{w_X(s')}{\rho_X}}
  w_X(s')}{\sum_{s'=1}^T e^{-\frac{w_X(s')}{\rho_X}}}
\end{eqnarray}

\section{Derivation of the maximum likelihood estimator for the Gumbel
  density for the minimum values}
\label{sec:derivation-min}
Take the log--likelihood function of the Gumbel density given by
Eq. (\ref{eq:Gumbel-min}) for $T$ observations $v_X(s') \quad
(s'=1,\ldots,T)$:
\begin{eqnarray}
\nonumber
l(\mu_X,\rho_X) &=& \sum_{s'=1}^T \ln\Bigl[
\frac{1}{\rho_X}\exp\Bigl(\frac{v_X(s')+\mu_X}{\rho_X}
-e^{\frac{v_X(s')+\mu_X}{\rho_X}}\Bigr) \Bigr] \\
\nonumber
&=& -T \ln \rho_X + \frac{1}{\rho_X}\sum_{s'=1}^T v_X(s') \\
&+& \frac{\mu_X}{\rho_X}T - \sum_{s'=1}^Te^{\frac{v_X(s')+\mu_X}{\rho_X}}.
\label{eq:ll-min}
\end{eqnarray}
Partially differentiating Eq. (\ref{eq:ll-min}) in terms of $\mu_X$ and
setting it to zero yields
\begin{eqnarray}
\nonumber
\frac{\partial l}{\partial \mu_X} &=& \frac{T}{\rho_X} -
\sum_{s'=1}^T\frac{1}{\rho_X} e~{\frac{v_X(s')+\mu_X}{\rho_X}} = 0,\\
\nonumber
T &=& \sum_{s'=1}^T e^{\frac{v_X(s')+\mu_X}{\rho_X}} \\
e^{-\frac{\mu_X}{\rho_X}} &=& \frac{1}{T} \sum_{s'=1}^T
e^{\frac{v_X(s')}{\rho_X}}.
\label{eq:eq-of-mu-min}
\end{eqnarray}
Similarly, differentiating Eq. (\ref{eq:ll-min}) in terms
of $\rho_X$ and setting it into zero yields
\begin{eqnarray}
\nonumber
\frac{\partial l}{\partial \rho_X} &=& -\frac{T}{\rho_X} - \frac{1}{\rho_X^2}
\sum_{s'=1}^T v_X(s') - \frac{\mu_X}{\rho_X^2}T \\
\nonumber
&-& \sum_{s'=1}^T
\Bigl(-\frac{v_X(s')+\mu_X}{\rho_X^2}e^{\frac{v_X(s')+\mu_X}{\rho_X}}\Bigr)
= 0, \\ 
\nonumber
-T\rho_X &-& \sum_{s'=1}^T v_X(s') - \mu_X T \\
\nonumber
&+& e^{\frac{v_X(s')}{\rho_X}}\sum_{s'=1}^T
(v_X(s')+\mu_X) e^{\frac{v_X(s')}{\rho_X}} = 0, \\
\nonumber
\rho_X + \mu_X &=& -\frac{1}{T}\sum_{s'=1}^T v_X(s') +
    e^{\frac{\mu_X}{\rho_X}} \frac{1}{T} \sum_{s'=1}^T v_X(s')
    e^{\frac{v_X(s')}{\rho_X}} \\
&+& e^{\frac{\mu_X}{\rho_X}}
    \frac{\mu_X}{T} \sum_{s'=1}^T e^{\frac{v_X(s')}{\rho_X}}.
\label{eq:eq-of-rho-min}
\end{eqnarray}
Inserting Eq. (\ref{eq:eq-of-mu-min}) into
Eq. (\ref{eq:eq-of-rho-min}) 
yields
\begin{eqnarray}
\nonumber
\rho_X + \mu_X &=& -\frac{1}{T}\sum_{s'=1}^T v_X(s')
+ \frac{\sum_{s'=1}^T v_X(s')
e^{\frac{v_X(s')}{\rho_X}}}{\sum_{s'=1}^T v_X(s')} + \mu_X \\
\rho_X &=& \frac{\sum_{s'=1}^T e^{\frac{v_X(s')}{\rho_X}}
  v_X(s')}{\sum_{s'=1}^T e^{\frac{v_X(s')}{\rho_X}}} 
- \frac{1}{T} \sum_{s'=1}^T v_X(s')
\end{eqnarray}

\end{document}